# USING PLANETLAB TO IMPLEMENT MULTICAST AT THE APPLICATION LEVEL

Genge Béla[1] and Haller Piroska[2]

[1] Department of Electrical Engineering, "Petru Maior" University of Tîrgu Mureș, Romania
`bgenge@engineering.upm.ro`

[2] Department of Electrical Engineering, "Petru Maior" University of Tîrgu Mureș, Romania
`phaller@engineering.upm.ro`

## ABSTRACT

*Application-layer multicast implements the multicast functionality at the application layer. The main goal of application-layer multicast is to construct and maintain efficient distribution structures between end-hosts. In this paper we focus on the implementation of an application-layer multicast network using PlanetLab. We observe that the total time required to measure network latency over TCP is influenced dramatically by the TCP connection time. We argue that end-host distribution is not only influenced by the quality of network links but also by the time required to make connections between nodes. We provide several solutions to decrease the total end-host distribution time.*

## 1. INTRODUCTION

For several years now group communications have been receiving significant attention from both the industry and scientific communities [1, 2]. One of the main applications of group communications is in the field of *multicast*. Historically speaking, the first multicast applications were implemented over the IP layer, also known as *IP multicast* [3]. However, after nearly a decade of research in the field of IP multicast, it was never fully adopted because of several technical and administrative issues [4].

Later, there have been several proposals for other multicast implementations that would be easier to deploy over the already existing and well-established Internet protocols and would require little or no modifications in existing routers. Such a survey of existing solutions was provided by El-Sayed et al [5].

One of the directions that has been clearly adopted over the last few years is *application-layer multicast*, which implements the multicast functionality at the application layer. The main goal of application-layer multicast is to construct and maintain efficient distribution structures between *end-hosts*. These structures are constructed using an *overlay* network providing the necessary infrastructure for data transfer between end-hosts.

Today's research focuses on the many aspects of application-layer multicast, including construction of overlay networks [6, 7], optimization issues [8] or security [9]. In our previous work [10] we have addressed the problem of optimally distributing end-hosts (i.e. EH) to overlay network hosts (i.e. OH) in order to minimize network latency and to distribute the load of OH. Based on a heuristic algorithm we proved that the algorithm ensures a local optimal distribution of EH in real time and thus can be used to provide a feasible solution to the distribution problem.

In this paper we focus on the actual deployment of the algorithm proposed in our previous work in a real and globally-scaled distributed system: *PlanetLab* [11]. *PlanetLab* is a "geographically distributed overlay network designed to support the deployment and evaluation of planetary-scale network services" [11]. Using PlanetLab, researchers can test their algorithms and systems in a real environment where nodes can become unreachable, network bandwidth can fluctuate and node processing capabilities can drop dramatically.

In order to test the real applicability of our previously proposed algorithm we have developed an overlay network in PlanetLab where nodes are connected in a complete graph model. There are several advantages for using such a graph model. First, there is no need for implementing complex routing algorithms [12, 22, 23], which greatly simplifies the implementation and functionality of the overlay. Second, maintaining routing tables is not more complex than maintaining connections with all the other nodes. As a downside of this topology, there are a large number of connections that must be maintained, which grows exponentially with the number of OH. However, the simplicity of the routing algorithms between OH makes this topology a great candidate for using it as a leaf component in hierarchical topologies [13, 14].

Existing research [6, 7, 15] focuses on measuring the delay between nodes after the overlay has been constructed or measuring the overlay construction time after TCP connections are done. In deploying our algorithm we have observed that the total time required to measure network latency over TCP is influenced dramatically by the TCP connection time. In this paper we also argue that end-host distribution is not only influenced by the quality of network links but also by the time required to make connections between nodes.

The paper is structured as follows. In Section 2 we provide a brief overview of PlanetLab, its concept and strengths. In Section 3 we provide an overall presentation of the overlay network, we discuss our previous work and we identify the main problems for deploying the previously proposed algorithm. In Section 4 we present the measurement results that were done with nodes spread across 23 countries and we provide 3 solutions for improving the performance of the measurements. Finally, we conclude with an overview of the proposed solutions and we mention some future solutions that could also be implemented.

## 2. PLANETLAB

In this section we briefly present the concept and architecture of PlanetLab [11]. Currently, PlanetLab serves as a networking test-bed for researchers running applications, protocols and algorithms. The PlanetLab project was launched in 2002 at the Princeton University, at that time it consisted of 100 machines distributed to 40 sites. Today, PlanetLab consists of 1133 nodes at 515 sites.

Its unexpected success was caused by many factors. First of all, nodes are using existing Internet connections. This means that from a connection point of view a new node only requires a public IP and an Internet connection. Also, as a basic requirement, nodes must not be protected by any firewalls or proxies, they must be publicly available. Such an approach reduces expenses providing at the same time a rapid growth of the test-bed.

A second factor that contributed to the success of PlanetLab was the node administration support. Administrating PlanetLab nodes for local site administrators is a simple task, as most of the tasks are covered by the central PlanetLab administration. From our experience, we have found this service extremely helpful, as it automatically handles package updates, kernel patches, and security issues. The only task of local administrator is to make sure that nodes are running and that they are connected to the Internet. This is also a key factor as for handling such a reduced set of tasks there is no need for hiring or training any personnel. This can simply be handled by the sites principal investigator (i.e. PI) or by any already existing staff member.

As PlanetLab was designed to be rapidly available for researchers it has a simple and efficient experiment set-up support, through a Web interface. By using this interface, the PI can create new experiments, it can add users to an experiment, and it can add any number of nodes to the experiment from the total of 1133 nodes. Access to nodes is made through SSH connections. Users are required to upload their SSH public key via the same Web interface. When a user is added to an experiment, his public key is broadcasted to all nodes in the experiment. The user is then authenticated using his private key.

Another aspect that increased the success of PlanetLab is that it provides users with bare Linux nodes, where they can upload and run any service. This is a clear advantage over the Grid, as it provides a much more flexible platform with the possibility of running a wide variety of protocols and services. A similar approach we found in Emulab [17] with the difference that PlanetLab nodes can connect and receive connections to and from nodes not belonging to PlanetLab. Also, an advantage of PlanetLab over Emulab is that PlanetLab nodes can run multiple separate experiments on the same node using virtualization, while Emulab nodes do not use virtualization techniques, thus nodes used in one experiment cannot be also used in another experiment.

As a final note on PlanetLab, we should mention that it has been used to evaluate a wide variety of network services such as content distribution [18, 19], large file distributions [20], measurement and analysis [21].

## 3. OVERLAY TOPOLOGY AND END-HOST DISTRIBUTION ALGORITHMS

The measurements that follow in the next sections are based on a complete graph overlay topology where EH are distributed using a heuristic algorithm. An example of such a topology is given in Fig. 1, where we have illustrated the presence of 3 host types:

- End-hosts (i.e. EH);
- Overlay-hosts (i.e. OH);
- Monitor-hosts (i.e. MH).

EHs are the producers and consumers of data transferred by the overlay. MHs are used to monitor the load of each OH and to distribute EHs to the least loaded OH. We identify two types of EHs: *measuring EHs* and *streaming EHs*. *Measuring EHs* denote EHs that connect to OHs in order to measure network latency. These measurements are then sent to the MH that runs the distribution algorithm presented in our previous work [10]. The MH then sends back the selected OH to which EHs connect and request resources. Through this last step, EHs become *streaming EHs*.

As mentioned before, the role of the MH is to distribute EHs to OHs. The algorithm we used in our previous work to distribute EHs relies on connection latencies measured by EHs to each OH (i.e. Alg. 1). These are then sent to the MH that chooses an OH such that the overall graph latency has a local optimal value. We use a local optimal value instead of a global one because from our simulation results this approach runs in the order of milliseconds, while the global optimal algorithm runs in the order of minutes for several thousand EHs and several hundred OHs.

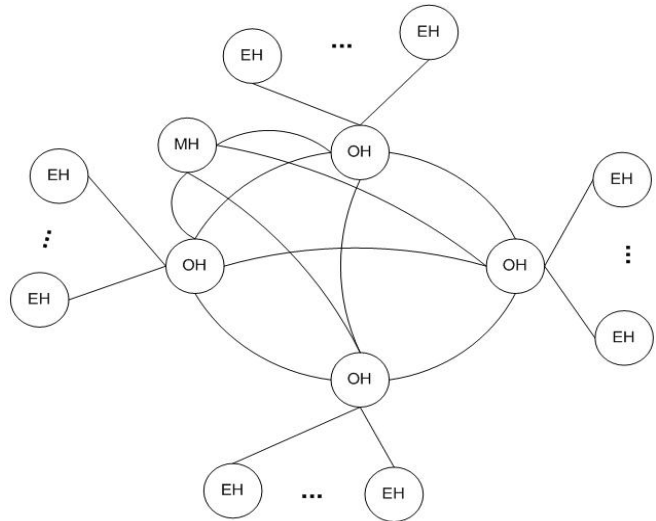

Figure 1. Multicast topology

---

**Algorithm 1** End-Host distribution algorithm

---

Let *Req* be the set of OH – measured latency pairs (*oh*, *l*)
Let $Chosen_{OH}$ be the set of chosen OHs
Let $eh_{OH} = OH_1$ be the OH chosen for this EH
Let $l_{min} = MAX\_VAL$ be the minimal computed latency

{Search for the OH that minimizes the overall graph latency}
**for all** (*oh*, *l*) $\in$ *Req* **do**
 Let $l = Latency(Chosen_{OH} \cup \{oh\})$
 **if** $l < l_{min}$ **then**
  $eh_{OH} = oh$
  $l_{min} = l$
 **end if**
**end for**

{Save the chosen OH for next EH distributions}
$Chosen_{OH} = Chosen_{OH} \cup \{eh_{OH}\}$

---

The distribution algorithm uses the measured latency between all OH pairs, the load of each OH and the measured latency between each EH and OH pairs. The algorithm is run by the MH each time a new EH must be connected. At this time, the EH must provide the MH its measurement results on the network latency it recorded to each OH. Based on this data and the reported load received from each OH, the MH runs the distribution algorithm.

We have chosen to deploy the proposed multicast in PlanetLab because it provides globally-available network services that can be used to run any application type that can run on a Linux OS. From the beginning of the implementation process we had to deal with several problems. First of all, network connections between PlanetLab nodes or even node CPUs can be heavily loaded, sometimes even leading to SYN_ACK timeouts for TCP connections. Second, nodes can be rebooted at anytime by PlanetLab Central coordinators in order to ensure a software update, for software maintenance or simply because of some hardware problems. These problems must be handled by the MH in order to ensure that EHs are not distributed to such nodes and that already distributed EH nodes are redistributed if necessary (i.e. on OH failure).

We also encountered several problems on the EHs side. The proposed algorithm heavily relies on the measurement data provided by EHs. This means that when joining the network, all EHs must first measure the latency with all OHs and then send this data to MH. The problem with this approach is that in some cases the response time from OHs is very long, in the order of seconds as shown in the next sections. This leads to an overall distribution time in the order of seconds or even minutes, which is unacceptable.

## 4. MEASUREMENT ISSUES AND SOLUTIONS

### 4.1. Overlay Construction Time

Although the construction of the overlay is done only once, we consider that measuring the construction time can provide useful perspective of the time required to re-construct the overlay in possible future developments. The constructing of the overlay network is not made instantly. In order to evaluate the performance and the general usability of the proposed overlay, we have measured the time needed to construct the complete graph between the overlay nodes.

Deploying and starting applications on PlanetLab nodes can be done automatically using applications such as *multicopy* or *multiquery* that are part of the CoDeploy project [16]. These allow a parallel deployment and execution of commands on a set of nodes. We have considered 5 settings with a different number of OH nodes. The OH applications were deployed on nodes from 14 countries (for the maximum number of 40 OH nodes), as shown in Table 1. After starting the OH applications each OH connects to all other OH according to Alg. 2, where `OH` corresponds to the set of OH, `Cout` is the set of outgoing connections and `Cin` is the set of incoming connections.

At first, each OH starts the connection process to other OH nodes. Then, it waits for the connection process to complete. This process leads to duplicate connections between each OH node pair. In order to eliminate duplicate connections we measure the connection latency in each direction by sending a single package of 1500Bytes and we eliminate the connection with the maximum latency.

Table 1. Country and OH node count

| Country | Node count | Country | Node count |
| --- | --- | --- | --- |
| Austria | 1 | Italy | 6 |
| Canada | 2 | Korea | 2 |
| France | 4 | Poland | 3 |
| Germany | 9 | Romania | 2 |
| Greece | 1 | Spain | 2 |
| Hungary | 1 | Switzerland | 1 |
| Israel | 1 | US | 5 |

**Algorithm 2** Constructing complete connections for one OH

---

   Let $t_1$ = @Get_curr_time()
   Let Cout = $\phi$

   {Start connection sequences}
   **for all** $oh \in$ OH **do**
     $c$ = @Start_conn_sequence($oh$)
     Cout = Cout $\cup$ {c}
   **end for**

   {Wait for completion}
   @Wait_for_completion( Cout )

   {Eliminate duplicate connections}
   Let Cin = @Get_incoming_connections()
   **for all** $c \in$ Cout **do**
     **if** $\exists c' \in$ Cin such that @Src_address($c'$) = @Dest_address($c$) **then**
        ($Meas_{out}$, $Meas_{in}$) = @Run_measurements($c$, $c'$)
        **if** $Meas_{out} < Meas_{in}$ **then**
           @End_connection($c$)
           Cout = Cout \ {c}
        **end if**
     **end if**
   **end for**

   {Calculate complete connection time}
   Let $t_2$ = @Get_curr_time()
   Let $G_{time} = t_2 - t_1$

---

According to Alg. 2, each OH calculates a complete connection time $G_{time}$. The complete graph construction time is the maximum of these values, as shown in Fig. 2. As we can see from this figure, the construction of the overlay is greatly influenced by the number of nodes. However, the variation is not linear because the overlay also depends on other factors such as the quality of network connections and the load of nodes. This result has the following explanation. In the first OH set (i.e. 3 nodes), all 3 nodes are located in European countries, with a minimum load. In the next OH set (i.e. 10 nodes) we have added additional nodes from Europe, one node from the US and one node from Asia. This almost doubled the graph construction time because the node from Asia was heavily loaded, with the CPU running at over 80% almost all the time. In the next set (i.e. 20 nodes) we have added additional nodes from Asia, Canada and Europe which, because of network connection latencies and heavily loaded nodes (i.e. from Israel and Germany) has led to a quadruple time. In the next two sets (i.e. 30 and 40 nodes) we have added additional nodes from Europe and US, leading to the results shown.

### 4.2. EH Connection Measurement Issues

When EH nodes are first started, each node connects to all OH nodes in order to measure the network latency. The measured values are then sent to the MH that applies Alg. 1 to determine the OH node where each EH must connect. We have identified two components that significantly influence the measured values: connection time and network latency. Let EH be the set of EHs. Then, the total measurement time $M_i$ needed to be executed by an EH is:

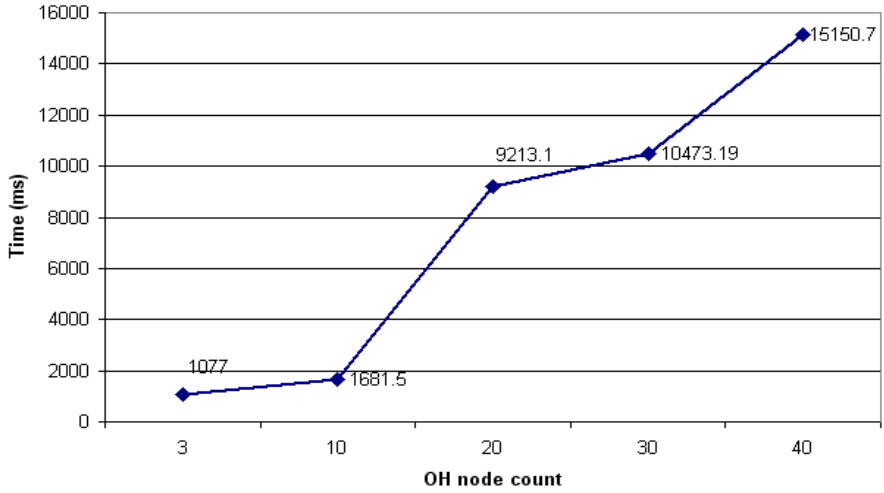

Figure 2. Complete graph construction time

$$M_i = \max_{oh_j} \{Conn(eh_i, oh_j) + CummLat(eh_i, oh_j)\},$$

$$CummLat(eh_i, oh_j) = Lat_1(eh_i, oh_j) + Lat_2(eh_i, oh_j) + Lat_3(eh_i, oh_j),$$

where $eh_i \in$ EH, $oh_j \in$ OH, $Conn$ denotes the time needed to establish a connection between $eh_i$ and $oh_j$ and the $Lat_x$ functions denote the round-trip latency of 3 packages.

We have considered several scenarios, with EHs count ranging from 10 to 1000. EHs were deployed on nodes from 23 countries (for the maximum number of 1000 EH nodes), as shown in Table 2.

Table 2. Country and EH node count

| Country | Node count | Country | Node count |
|---|---|---|---|
| Argentina | 10 | Japan | 10 |
| Australia | 10 | Korea | 20 |
| Austria | 40 | Netherlands | 20 |
| Belgium | 20 | Poland | 40 |
| Canada | 100 | Portugal | 10 |
| China | 20 | Romania | 20 |
| Finland | 10 | Russia | 20 |
| France | 110 | Spain | 40 |
| Germany | 160 | Switzerland | 10 |
| Greece | 10 | Taiwan | 10 |
| Hungary | 20 | US | 240 |
| Italy | 60 | - | - |

Each EH calculates its own $M_i$ value that is sent to the MH that calculates an average measurement time, illustrated in Fig. 3. We can see that the number of OH nodes clearly influences the overall measurement time. There are several values that break the linear trajectory. For instance, in the case of 40 OH nodes, when running 50 EH nodes the average time is 39382ms and when running 100 EH nodes the average time is reduced to 21571ms. The explanation for this behaviour lies in the way that the measurements were done. Because PlanetLab offers a set of resources over the Internet that is shared among researchers, time measurements can change dramatically from one execution to another. Moreover, the measurements we made span across 10 days. We have actually seen that in one day a given node can be extremely loaded because other researchers may also be running experiments, and the next day the node can show a minimum load. This is in fact the expected behaviour of nodes running in a real networking environment that greatly differs from the controlled laboratory environments.

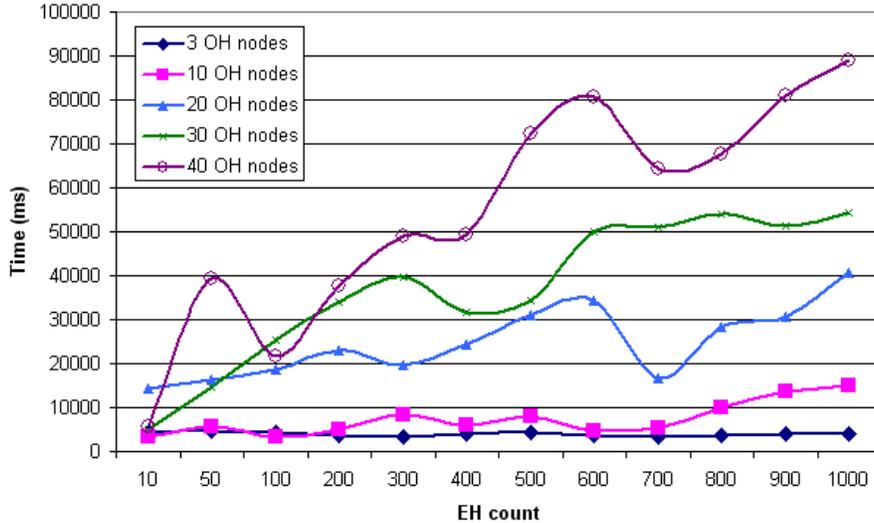

Figure 3. Average EH measurement time

The values shown in Fig. 3 include both the connection time and the network latency. However, as shown in Fig. 1 the latency is only a small part of the measurement time, with average values ranging from 68.59ms to 925.86ms.

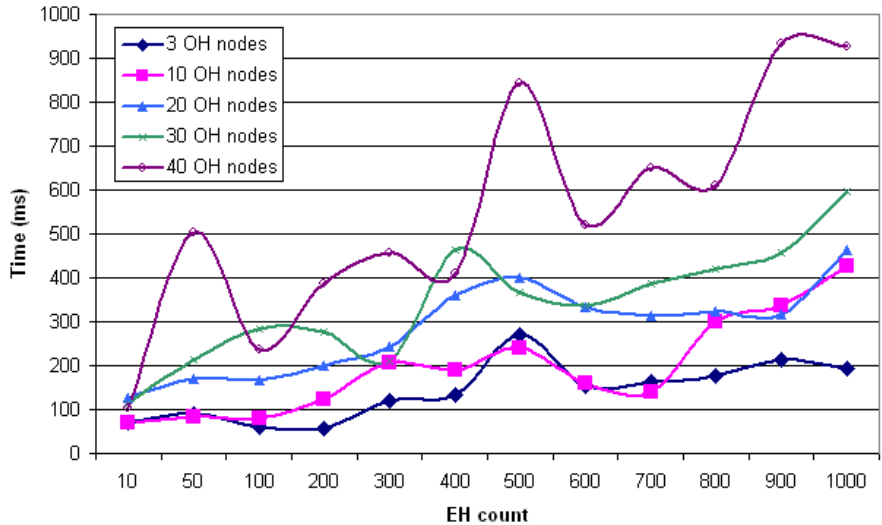

Figure 4. Average EH-OH measured latency

The values shown in Fig. 3 clearly show that we should improve the performance of the measuring algorithm. At this stage, the average time needed to measure the network latency for 1000 EH nodes in the 40 OH nodes setting is 89000ms, which corresponds to almost 1.5 minutes. However, this is the average time, which is much smaller than the maximum time needed for an EH to make the measurements. The maximum measurement time is shown in Fig. 5, where we can see that the maximum time needed to make the measurements is in fact 561192ms, which is almost 9.5 minutes. These values clearly show that the time needed for all nodes to make the measurements is influenced by the number of OHs and by the number of EHs, leading to the value of 9.5 minutes, which is unacceptable.

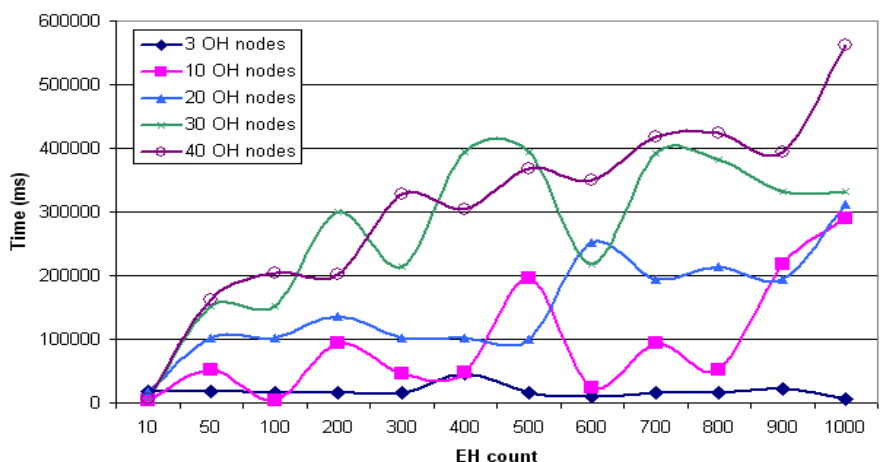

Figure 5. Maximum EH measurement time

The total accessing distribution time of EHs is also influenced by the response time from the MH. In all our measurements the MH resides on a single node from Romania. In Fig. 6 we can see the average response time from the MH. Interestingly, the response time is not influenced by the number of OHs or by the number of EHs, but by the number of simultaneous requests that are received. EHs connect to the MH only after completing the measurements; this is why when a large number of EHs connect simultaneously to the MH we get the peaks from the figure. From the measurements we have also seen that after receiving the measurement data the distribution algorithm is running under 1ms for each request, thus the values shown in Fig. 6 are given by message processing and network delay.

After an EH successfully connects to the OH, it can stay connected for an unlimited time. However, if the connection is interrupted, it will reconnect to the designated OH. If the designated OH is no longer available, it must execute the measurement and distribution all over again. In case of new EH nodes, these are distributed by the MH without redistributing the already connected EH nodes.

As mentioned earlier, in case of OH failure, disconnected EH nodes initiate a new measurement and distribution process. However, in case of network failures between OH nodes, a reconnect mechanism is activated for each OH node that tries to re-establish connection with all other OH nodes, effectively trying to reconstruct the overlay.

### 4.3. EH Connection Measurement Solutions

As illustrated in the previous section, making network measurements at the application layer is mainly influenced by the connection time between nodes. The network latency factor, as opposed to the connection time, has a minimum impact on the total time.

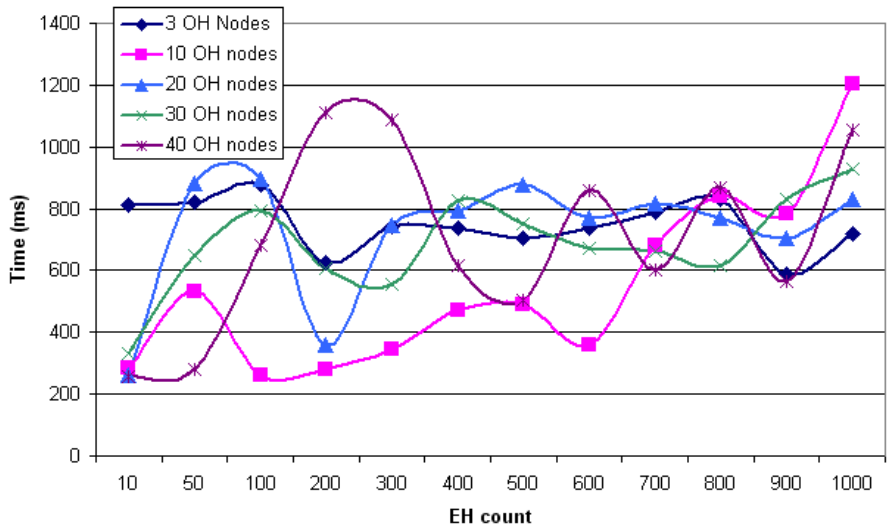
Figure 6. Average MH response time

When EHs use the proposed overlay, their main goal is not to make measurements but to actually use it to effectively distribute data. The time needed to make the measurements should thus be reduced to a minimum possible.

In this section we propose 3 solutions to the measurement problem. After implementing them, we have repeated the measurements for the 1000 EH setup, where the modifications would have a greater impact.

The first solution involves reducing the reconnect process count to 0, meaning that if a connect attempt fails, the EH removes the OH from its list. EH nodes usually try to connect over and over again to OH nodes until successful. This process dramatically increases the overall measurement time, as shown in the previous section. By eliminating the reconnections, we are in fact eliminating OHs that are overloaded or to which we have a poor connection. The improvements can be immediately seen, as shown in Fig. 7. In this case, for the maximum setting, with 40 OH nodes, the average measurement time drops from 89000ms to 22027ms, improving the overall measurement 4 times.

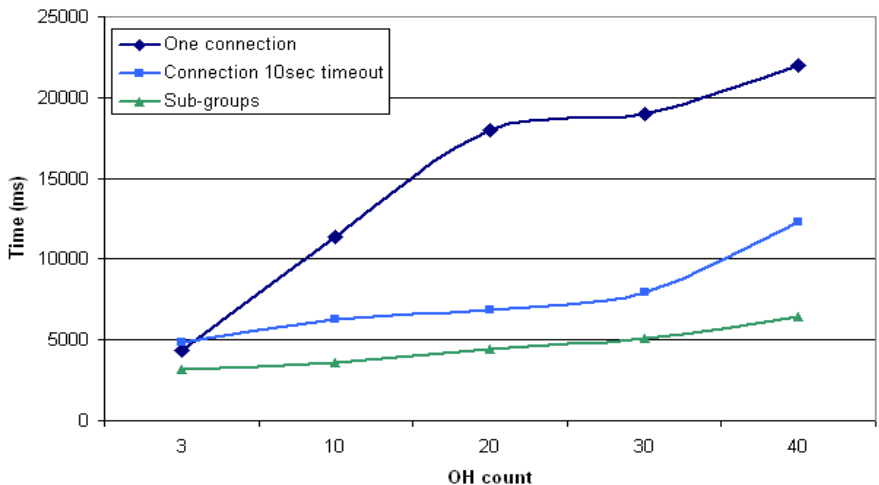
Figure 7. Average improved EH measurement time for 1000 EH

The problem with the first solution is that a connection must be timed out by the Operating System (i.e. OS) to eliminate the OH from the solution. As a second solution we propose an application-controlled connection timeout, opposed to network OS timeout. In this case we timed out connections that exceeded 10 seconds, decreasing the average measurement time from 89000ms to 12284ms and improving the overall measurement 7 times, as shown in Fig. 7. The 10 seconds were chosen based on the observation that a lower timeout leads to an increased number of OH nodes eliminated from the solution. This problem is discussed in more detail later in this section.

The third solution involves partitioning the OH and EH nodes into sub-groups, thus reducing the total number of OHs/EHs and the total number of EHs/OHs. The partitioning can be seen in Table 3. As shown in Fig. 7 the average time required for measurements is reduced to 6459ms for 40 OH nodes, improving the overall measurement time over 13 times.

Table 3. Sub-group partitioning

| Sub-Group | 3 OHs 1 OH/EH | 10 OHs 2 OH/EH | 20 OHs 4 OH/EH | 30 OHs 6 OH/EH | 40 OHs 8 OH/EH |
|---|---|---|---|---|---|
| Grp1 | 333 EH | 200 EH | 200 EH | 200 EH | 200 EH |
| Grp2 | 333 EH | 200 EH | 200 EH | 200 EH | 200 EH |
| Grp3 | 333 EH | 200 EH | 200 EH | 200 EH | 200 EH |
| Grp4 | - | 200 EH | 200 EH | 200 EH | 200 EH |
| Grp5 | - | 200 EH | 200 EH | 200 EH | 200 EH |

The direct effect of the first two solutions is that the number of OHs for which EHs test the connection reduces significantly with the reduction of timeouts. For instance, by using the OS timeout, ranging from a few seconds to a few minutes, we have less eliminated OHs than using a fixed timeout of 10 seconds, as shown in Fig. 8. In case of only one connection (i.e. OS timeout) the tested percentage is 100% for 3 OHs, however, this drops to 95% for 10 and 20 nodes and then rises to 96.66% for 30 nodes and to 97.43% for 40 nodes. In case of application-layer timeout we have 98.1% for 3 OHs which drops to 71.79% for 40 OHs.

Although the partitioning-based solution provides the best timings, it can limit sub-groups to a set of OH nodes that may not provide the optimal solution for the entire group. While the application-layer timeout mechanism seems to be the next best approach, care must be taken in choosing the timeout value because a larger connection-time does not necessarily mean that the specific node is heavily loaded, but several other factors can also influence this value, such as a momentarily busy OS, or a momentarily busy application.

Other solutions could also be applied, such as using UDP for determining the network latency between EHs and OHs. Such a solution would eliminate the overhead given by TCP connection. However, because the overlay uses TCP for forwarding data, making measurements by connecting to OHs via TCP provides a more precise view on the future behaviour of OH nodes.

## 5. CONCLUSIONS AND FUTURE WORK

We presented several issues and solutions for deploying application-layer overlay networks. Based on our measurements conducted over PlanetLab, a real network testing platform, we have concluded that distributing EHs cannot be based only on the measured network latency, but must also include other elements such as connection time or EH geographical location to reduce the time required to make the actual latency measurements.

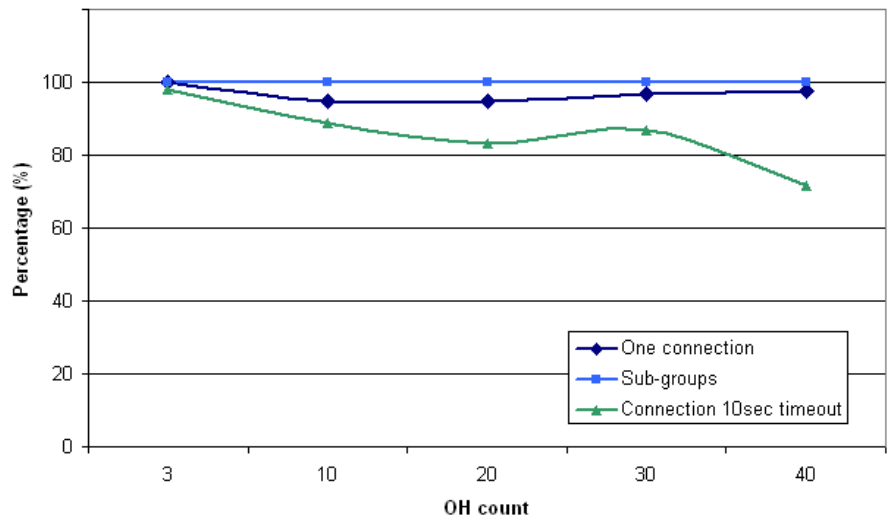

Figure 8. Average percentage of connections measured

The identified problems have several solutions. In this paper we have proposed 3 such solutions: a first one that eliminates reconnections, a second one that uses application-layer timeouts and a third one that constructs sub-groups for reducing the number of OHs/EHs and EHs/OHs. By using these solutions we have shown that the measurement time can be reduced up to 13 times for 1000 EHs and 40 OHs.

As future work, we intend to use UDP for the initial measurements. However, special care must be taken because a lower timing for UDP packages does not necessarily imply lower timings for TCP packages. A study must be made to determine the correspondence between UDP and TCP timings and how could UDP-based measurements be used to forecast the overhead introduced by TCP connections. This study must also take into consideration UDP packet losses that may also influence the total measurement time.

## Authors

Genge Béla is an Assistant Lecturer at the "Petru Maior" University of Tîrgu Mure , Romania and a Consultant for Intelligent Building Solutions, Tîrgu Mure , Romania. In 2009 he received his PhD in Computer Science from the Technical University of Cluj Napoca, Romania in the field of network security protocol composition. His main research activities are related to network security, security protocols, and multimedia systems. He is also a reviewer for several international journals and conferences.

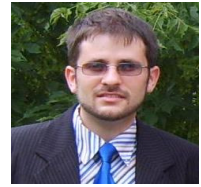

Haller Piroska is an Associate Professor at the "Petru Maior" University of Tîrgu Mure , Romania and a Consultant at Intelligent Building Solutions, Tîrgu Mure , Romania. In 2002 she received her PhD in Computer Science from the Technical University of Cluj Napoca, Romania in the field of distributed multimedia systems. Her main research activities are related to the data transfer optimization.

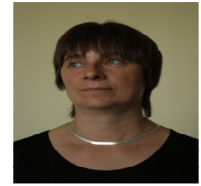